\documentclass[prb,twocolumn,aps,superscriptaddress,showpacs]{revtex4-2}

\usepackage{hyperref}
\usepackage{graphicx}
\usepackage{amsmath}
\usepackage{physics}
\usepackage{amsfonts}
\usepackage{amssymb}
\usepackage{bm}
\usepackage[usenames,dvipsnames]{color}
\usepackage{float}
\usepackage{natbib}
\usepackage{comment}
\hypersetup{
	colorlinks=true,
	citecolor=blue,
	linkcolor=blue,
	urlcolor=blue}

\begin{document}
	\newcommand{\be}{\begin{equation}}
	\newcommand{\ee}{\end{equation}}
	\newcommand{\bea}{\begin{eqnarray}}
	\newcommand{\eea}{\end{eqnarray}}
	\newcommand{\bec}{\begin{center}}
		\newcommand{\eec}{\end{center}}

	\title{Non-linear response of interacting bosons in a quasiperiodic potential}
	
	\author{Debamalya Dutta}
	\address{National Institute of Science Education and Research, Jatni, Odisha 752050, India}
	\address{Homi Bhabha National Institute, Training School Complex, Anushakti Nagar, Mumbai 400094, India}
	\author{Arko Roy}
	\address{School of Basic Sciences, Indian Institute of Technology Mandi, Mandi-175075 (H.P.), India}
	\address{INO-CNR BEC Center and Universit\`a di Trento, via Sommarive 14, I-38123 Trento, Italy}
	\author{Kush Saha}
	\address{National Institute of Science Education and Research, Jatni, Odisha 752050, India}
	\address{Homi Bhabha National Institute, Training School Complex, Anushakti Nagar, Mumbai 400094, India}

	\date{\today}

	\begin{abstract}
		We theoretically study the electric pulse-driven non-linear response of interacting bosons loaded in an optical lattice in the presence of an incommensurate superlattice potential. In the non-interacting limit $(U=0)$, the model admits both localized and delocalized phases depending on the strength of the incommensurate potential $V_0$. We show that the particle current contains only odd harmonics in the delocalized phase in contrast to the localised phase where both even and odd harmonics are identified. The relative magnitudes of these even and odd harmonics and sharpness of the peaks can be tuned by varying frequency and the number of cycles of the applied pulse, respectively. In the presence of repulsive interactions, the amplitudes of the even and odd harmonics further depend on the relative strengths of the interaction $U$ and the potential $V_0$. We illustrate that the disorder and interaction-induced phases can be distinguished and characterized through the particle current. Finally, we  discuss the dynamics of field induced excitation responsible for exhibiting higher harmonics in the current spectrum. 
		
	\end{abstract}
	
	\maketitle
	
	\section{\label{sec:level1}Introduction}
	The unprecedented controllability of ultracold gases offers a unique test bed for verifying several condensed matter phenomena ranging from the physics of non-interacting electrons to the physics of highly correlated electrons. For example, the celebrated single-particle Anderson localization\cite{anderson_1958} of non-interacting electrons can be realized \cite{expt1_Al_2008,expt2_Al_2008,expt3D_Al_2011,expt3D_Al_2012,white_20} in ultracold settings, whereas this phenomenon is difficult to observe in real materials due to suppression of disorder effect by a number of quantum phenomena\cite{FALLANI2008119}. The state-of-the-art ultracold atom experiments allow one to tune the atom-atom interactions to negligible value and to observe the single-atom behavior under the influence of disordered potential. This has motivated a great volume of works on ultracold bosons in the presence of disorder and weak interactions, revealing a plethora of intriguing collective localization phenomena\cite{chalker_PRL08,lugan_PRA11,lellouch_PRA15,lugan_PRL07,FALLANI2008119}. Moreover, the experimental feasibility to generate quasi-periodic optical potentials presents an ideal platform to investigate another paradigmatic localization, namely Andre-Aubry localization\cite{aubre_andre_1980} which shows localization-delocalization transition as the strength of the quasiperiodic potential is varied.

	Since the atom-atom interaction can easily be tuned to strong-coupling limit using an optical lattice potential, the study of the interplay between interaction and random or quasi-periodic disorder has received much attention in recent times\cite{yoo_PRB20}. It has been shown that the interplay between disorder (random or quasiperiodic) and interaction leads to many-body localized (MBL) states in the highly-excited spectrum\cite{basko_AOP06,huse_PRB13,pal_PRB10,potter_PRX15,serbyn_PRX15}. These many-body localized states fail to thermalize and cannot be described by the conventional statistical mechanics. It is now not a mere theoretical concept, rather a reality following an experimental evidence of the many-body localized state in a fermionic cold atomic setting\cite{borida_NatPhys17}. Furthermore, it has been shown that the interacting bosons in the presence of both random and quasiperiodic disorder exhibit a compressible insulating phase, namely Bose glass phase\cite{fisher_PRB89,giamarchi_PRA08,yao_PRL20}. In addition, very recently the experimentally realizable quasiperiodic bosonic model has been shown to exhibit MBL-ergodic phase transition\cite{anirban_PRB18}.
	Despite several studies, the interplay between disorder and interacting bosons and fermions remains an active area of research  towards investigating unconventional phases such as appearance of singular-continuous spectra, small interaction driven instabilities, anomalous transport, etc\cite{yevgeny_EPL17,luitz_PRB15,kohlert_PRL19,yoo_PRB20}.

	While there are extensive studies on revealing atypical localized phases in an interacting system with quasiperiodicity at equilibrium, the response of this system to an external field has not received much attention particularly in the non-linear regime. It is yet to be understood how different phases respond to the application of an external strong field. The reason for focusing on this particular dynamical aspect is attributed to the recent advancement of non-linear spectroscopy stemming from the matter-light interaction which can decode the microscopic properties of interacting systems. Although this is a decades-old field and widely studied in gaseous medium\cite{ferray_JPhysB1988,huillier_PRL1991,jeffrey_PRA1992,krause_PRL1992,macklin_PRL1993,huillier_PRL1993,lewenstein_PRA1994}, recent experimental realization of matter-light interaction in solid state systems\cite{ghimire_NatPhys11,ghimire_PRA12} has renewed interest to study matter-light interaction in various quantum systems due to potential application in attosecond science. Such systems include non-interacting Bloch solids\cite{ghimire_NatPhys19,yu_APX19,mengxi_PRA15}, Mott insulators\cite{silva_NatPhot18,murakami_PRB18,murakami_PRB21,arXiv:2203.01029}, Dirac insulators\cite{cheng_PRL20,taya_PRB21},twisted bilayer graphene\cite{ikeda_PRR20}, graphene\cite{mrudul_PRB21}, quantum spin liquids\cite{masahiro_PRR21}, quantum spin systems\cite{ikeda_PRB19}, etc. In addition, two of the authors of the present article have recently shown that the particle current in an interacting bosonic system can contain multiple odd harmonics of the applied field \cite{roy_RRR20} similar to real materials. 
	 
	 Partly enticed by the generation of higher harmonics in our previous study on interacting bosons under synthetic electric field, and the availability of experimentally realizable quasiperiodic potential in optical lattice settings, we address here how the non-linear response of interacting bosons to an electric pulse  gets affected once we introduce quasiperiodicity. We note that recently the field driven non-linear response has been studied in an {\it noninteracting} fermionic model in the presence of weak lattice potential involving disorder \cite{chinzei_RRR20} and quasiperiodic potential \cite{adhip_arXiv21}, however the interplay between interaction and quasiperiodicity in a bosonic model is yet to be addressed. We show that quasiperiodicity has dramatic effects on the non-linear response of the different equilibrium phases of the interacting bosonic model. In the non-interacting limit, the delocalized phase exhibits only odd harmonics in contrast to the localized phase where both even and odd harmonics are illustrated. For stronger localization, the magnitude of maximum harmonic orders (i.e., cutoff) reduces for a fixed pulse frequency due to the presence of large minigaps in the system. Remarkably, we find that in the localized phase even harmonic can be tuned by varying frequency. However, these features are absent in the delocalized phase. In the presence of interaction, the response of the field turns out to differ in the localized phase driven by interaction(Mott localization) from that of the localization due to quasiperiodicity (Aubry-Andre localization). Thus, the non-linear response may help in distinguishing and characterizing these two types of localization phenomena. Further we investigate the dynamics of excitations responsible for the emergence of multiple harmonics in the system.

	\section{\label{sec:level2}Model Hamiltonian\protect}
	The tight binding, time-independent Hamiltonian describing a 
	system of one-dimensional interacting bosons loaded in a quasi-periodic potential is given by
	\bea
	\hat{H}&=&-|J|\sum_jc_j^{\dagger}c_{j+1}+{\rm h.c.}+\frac{U}{2}\sum_j n_j(n_j-1)\nonumber\\
	&+&V_0\sum_{j}\cos{(2\pi\alpha j)}c_j^{\dagger}c_j,
	\label{tidh}
	\eea
	with $|J|$ as the hopping parameter, $U>0$ being the on-site repulsive interaction strength between the atoms, $V_0$ as the strength of the onsite potential, and $\alpha=(\sqrt{5}-1)/2$ being an irrational number. The bosonic creation(annihilation) operator are given by $c_j^{\dagger}(c_j)$ and $n_j = c_j^{\dagger}c_j$ is the number operator. 
	With an external driving via an electromagnetic field, the electric field $E(t)=-\partial_t A(t)$ couples synthetically to the neutral atoms
	through the time-varying vector potential $A(t)$. In particular,
	the tunneling term $J$ becomes complex with the Peierls phase.
	The effective time-dependent Hamiltonian in the velocity gauge
	assumes the form
	\bea
	\hat{H}(t)&=&-J(t)\sum_jc_j^{\dagger}c_{j+1}+{\rm h.c.}+\frac{U}{2}\sum_j n_j(n_j-1)\nonumber\\
	&+&V_0\sum_{j}\cos{(2\pi\alpha j)}n_j,
	\label{tdh}
	\eea
	where, $J(t)\equiv |J|e^{\iota \Phi(t)}$ with $\Phi(t)=qA(t)a/\hbar$, where $a$ is the lattice parameter, and $q$ is the effective
	charge. For the current work, we use a $n$-cycle $\sin^2$ time
	varying potential of the form of a pulse
	$A(t)=A_0\sin^2{(\omega t/2\,n)}\sin{(\omega\, t)}$ with $\omega$ being the frequency of oscillation ($\omega=2\pi n_0$). The strength of the vector potential $A_0\sin^2{(\omega t/2\,n)}$ smoothly varies with $t$ and the maximum value is attained at the half-cycle of the pulse. For rest of the work we measure $A_0$ in dimensionless unit and $n_0$ in THz. It is worth mentioning here that the dynamics of harmonic generation indeed depends on the shape of the pulse as discussed in Ref.~\cite{neyra_PRA21}.

		\begin{figure} 
		\includegraphics*[width=0.95\linewidth]{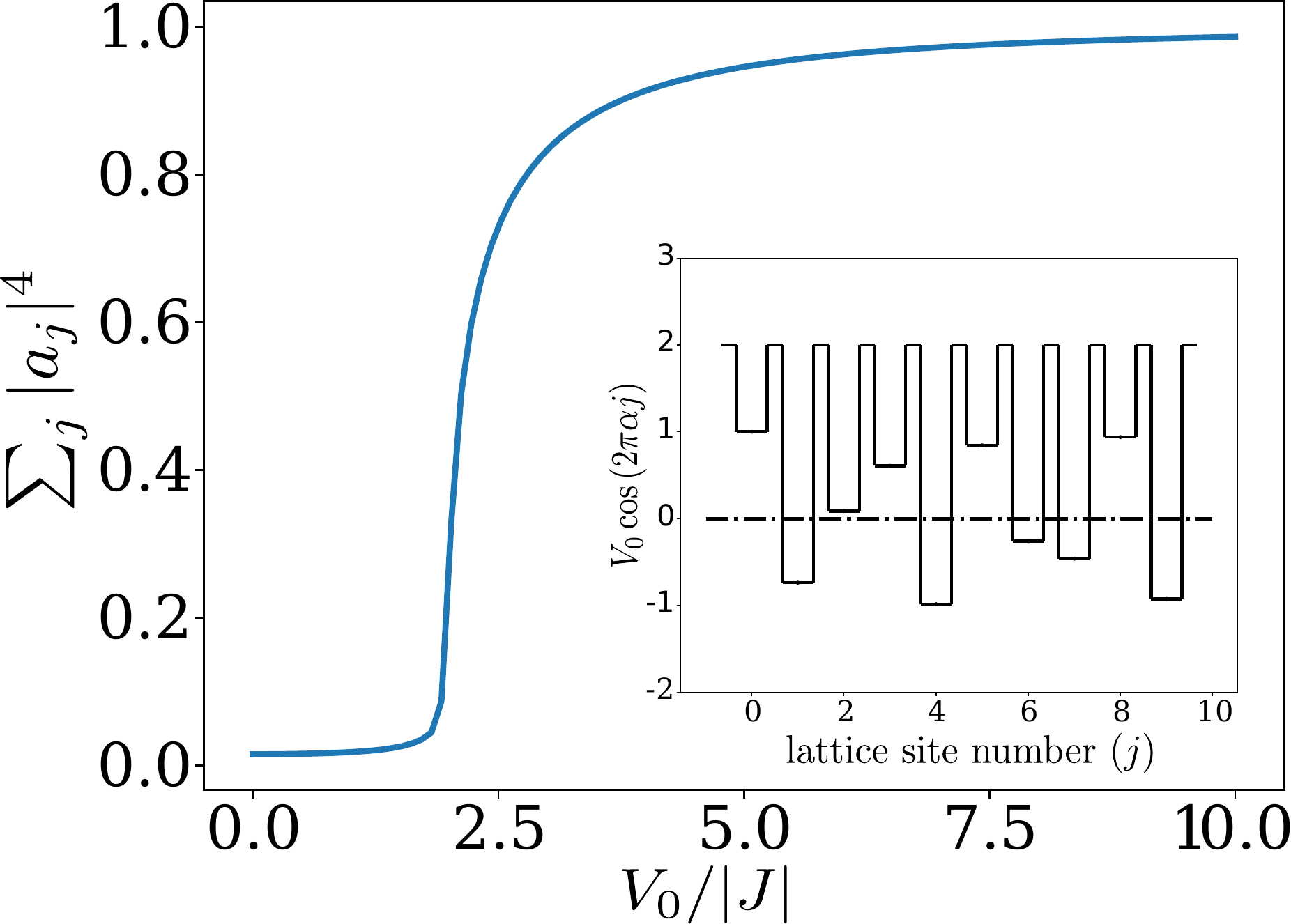}
		\caption{Variation of IPR with scaled onsite-potential($V_0/|J|$) diagram for non-interacting case($U=0$). (Inset) A schematic figure of the quasi-periodic potential with lattice site number $j$ has been provided in inset for $V_0=1$ (in units of energy).}
		\label{fig:2D_IPR}
	\end{figure}   
	\begin{figure*}[htbp]
		\includegraphics[trim={0cm 0cm 0cm 0cm},clip,width=0.95\linewidth]{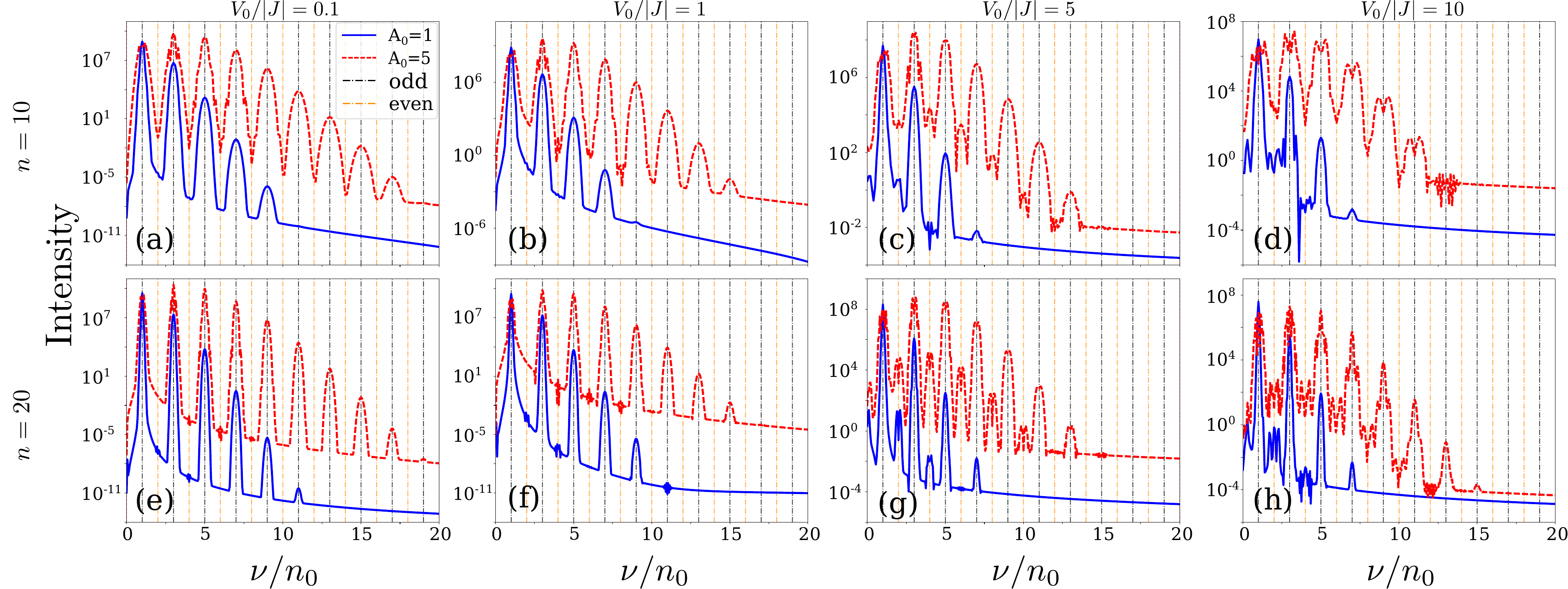}
		\caption{Plots showing intensity spectra with the multiplicity of incident frequency, for number of particle $(N)=1$ and number of lattice sites $(L)=200$ with strength of disordered potential $V_0/|J|$ and the number of cycles $n$ as parameter.}
		\label{fig:HHG_nonint}
	\end{figure*}

	\section{\label{sec:level6}Particle Current}
	The response of the external time-dependent electric field is computed by employing the current operator given by 
	\be
	\mathcal{\hat{J}}(t)=-i \frac{aq|J|}{\hbar} \sum_j(e^{i \Phi(t)}c_j^{\dagger}c_{j+1}-{\rm h.c.}).
	\label{current}
	\ee
	We then calculate the expectation of $\mathcal{\hat{J}}(t)$ with respect to the time evolved ground state $|\Psi_0(t)\rangle$ of the Hamiltonian, i.e. $\langle \mathcal{\hat{J}}(t)\rangle=\langle\Psi_0(t)|\mathcal{\hat{J}}(t)|\Psi_0(t)\rangle$.
	To find $|\Psi_0(t)\rangle$, we numerically solve the time-dependent Schr\"odinger equation
	$\hat H(t)\,\psi(t)= i\hbar \frac{\partial \psi(t)}{\partial t}$. For non-interacting Hamiltonian ($U=0$), we use single particle basis to construct the Hamiltonian for system size $L=200$ and subsequently diagonalize it to find ground state at $t=0$. In contrast, for $U\ne0$, the Hamiltonian is expressed in many-particle basis, and is restricted to lattice sites of length $L=7$ and total number of atoms $N=7$. The dimension of the Hilbert space increases exponentially with the increase in the system size in the bosonic model, and thus computing the dynamics becomes computationally expensive. In the current work, the ground state of the interacting Hamiltonian ($U\ne0$) at $t=0$ is computed by exact diagonalization. We then use fourth order Runge-Kutta algorithm for an optimum temporal step size which renders the dynamics convergent, to evolve $|\Psi_0(0)\rangle$ under the effect of time-dependent Hamiltonian $\hat H(t)$ to find $|\Psi_0(t)\rangle$. With these considerations, the modulus square of the Fourier transform of
	$\langle \mathcal{\dot{\hat{J}}}(t)\rangle$ (the rate of change of $\langle \mathcal{\hat{J}}(t)\rangle$ with time) provides information about the intensities and frequencies $\nu$ of non-linear excitations developed in this dynamical process.
	We next move on to demonstrate the effects of the time-dependent electric pulse field on the non-interacting as well as the interacting Aubry-Andre model. 
	\begin{figure}
		\includegraphics[width=0.95\linewidth]{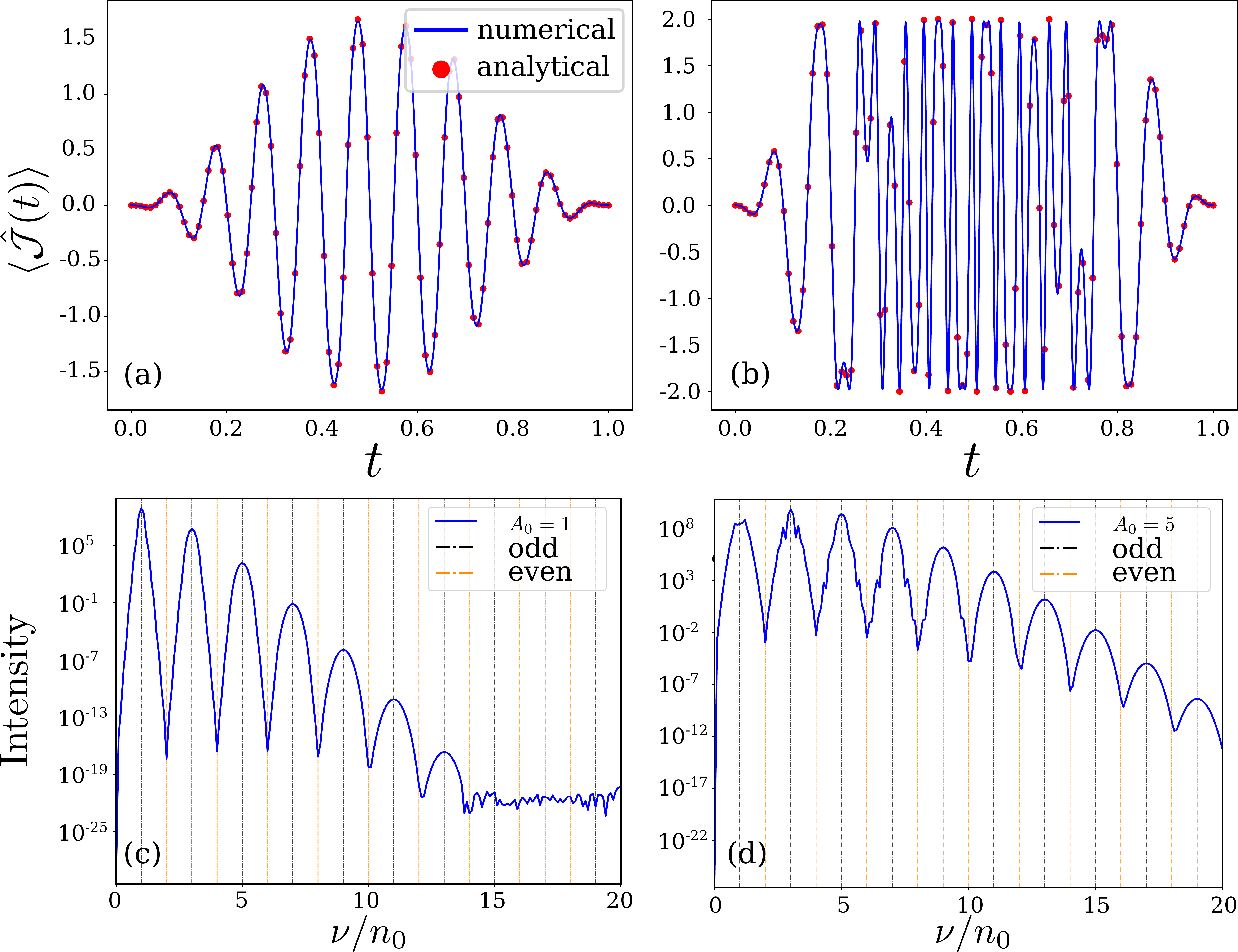}
		\caption{(a) and (b) represent variation of  $\mathcal{\hat{J}}(t)$ with time for noninteracting Bose-Hubbard model for $A_0=1$ and $A_0=5$ respectively. Here $t$ is
        measured in picoseconds. (c) and (d) represent the modulus square of Fourier transform of $\langle \mathcal{\dot{\hat{J}}}(t)\rangle$. The blue solid line (red dots) shows the numerical (analytical) results.}
		\label{fig:one_band_a-n}
	\end{figure}
				
	\section{Results}
	\subsection{Non interacting case ($U=0$)}
	In the non-interacting limit $(U=0)$, the model described in Eq.~(\ref{tidh}) admits 
	delocalized (localized) phase when $V_0/|J|<2 (V_0/|J|>2)$. Which
	is evident from Fig.~\ref{fig:2D_IPR} showing the variation of inverse participation ratio (IPR) with the relative strength of the disordered potential. On diagonalizing Eq.~\ref{tidh}, the ground state can be written as $|\Psi_0\rangle=\sum_j a_j|j\rangle$, where $|j\rangle$ is the site-basis and $a_j$'s are the coefficients of expansion. The IPR is then defined to be $\sum_j |a_j|^4$. 
	
	\begin{figure}
		\includegraphics[trim={0cm 0cm 0cm 0cm},clip,width=0.95\linewidth]{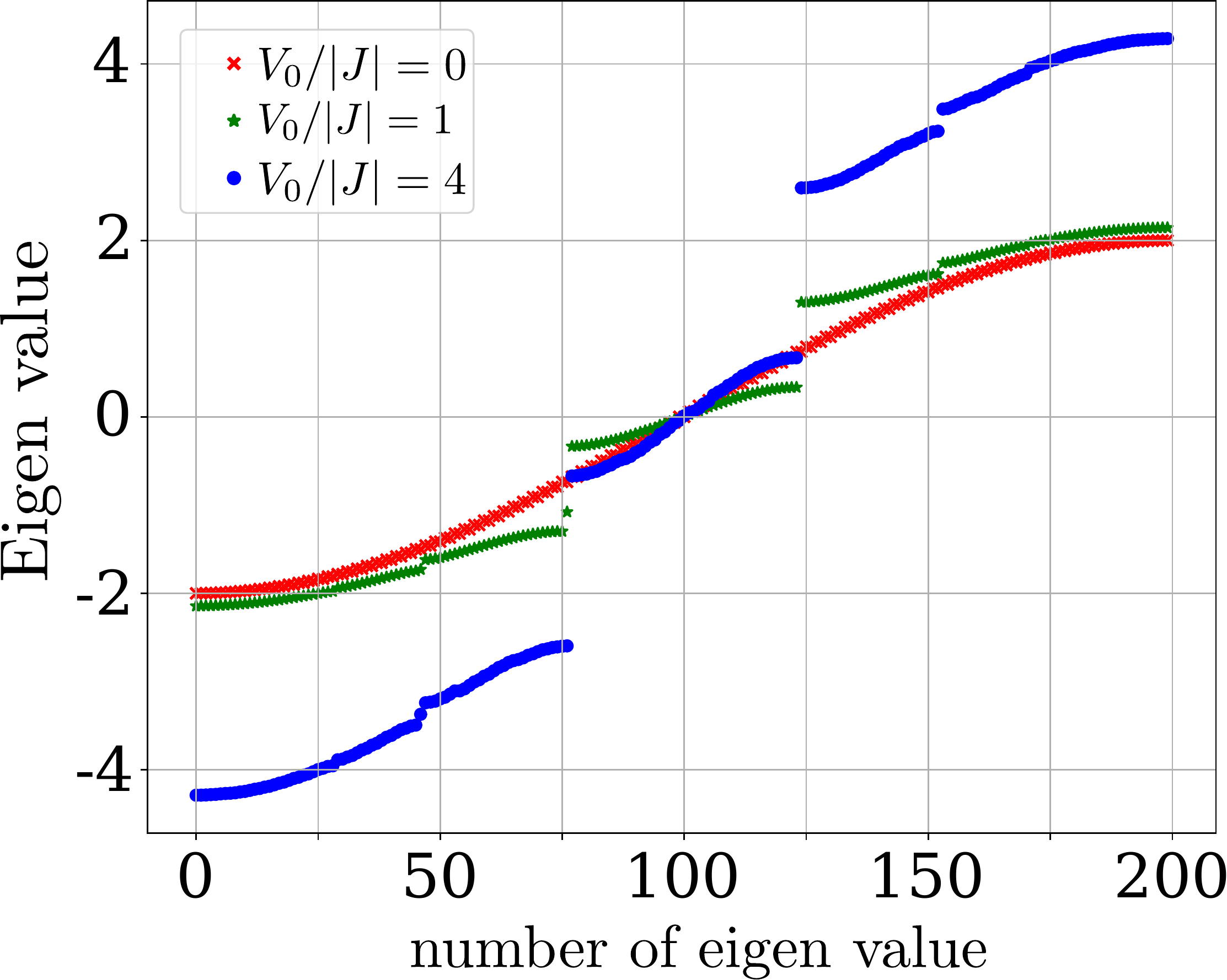}
		\caption{Band structure of the non interacting system for different strengths of disordered potential.}
		\label{fig:band_structure}
	\end{figure}

	\subsubsection{Intensity spectra and Mechanism}
	In the presence of $\sin^2$ pulse, the delocalized phase is identified by the appearance of only odd harmonics in the intensity spectrum of the response (see Fig.~(\ref{fig:HHG_nonint}a)) although the inversion symmetry of underlying Hamiltonian is broken. Notice that the order of harmonics increases with the increase in the applied field strength. The appearance of the intensity spectra with higher multiplicity of the incident frequency in the delocalized phase with $V_0<2|J|$ can be understood from the single-band physics with $V_0=0$. The intraband current for a single-band is given by $J_{\rm intra}=n_dqv_g$, where $v_g$ is the group velocity of the particle and $n_d$ is the particle density. The $v_g$ is computed as $v_g= {\partial \epsilon(k)}/{\partial k} = 2|J|a\sin{(ka)}$, where $\epsilon(k)=-2|J|\cos{(ka)}$ is the single-particle energy dispersion of Eq.~(\ref{tidh}) with $U=0$ and $V_0=0$. It is to be noted in the velocity gauge, due to driving, the crystal momentum $k$ becomes time-dependent and gets modified to $k_0 + q A(t)$. Together with $A(t)$ and $k$ we obtain 
	\bea
	v_g (t)&=&2a|J|\bigg[\sin{(k_0a)}\cos{\left\{qaA_0\sin^2{\left(\frac{\omega t}{2n}\right)}\sin{(\omega t)}\right\}}\nonumber\\
	&+&\cos{(k_0a)}\sin{\left\{qaA_0\sin^2{\left(\frac{\omega t}{2n}\right)}\sin{(\omega t)}\right\}}\bigg].
	\label{eq:group_velocity}
	\eea
	Fig.~(\ref{fig:one_band_a-n}) (a-b) illustrate $J_{\rm intra}(t)$ for different strength $A_0$ of the applied field computed using Eq.~\ref{eq:group_velocity}. Clearly, the intensity spectra $|\dot J_{\rm intra}(\nu)|^2$ contains higher harmonics of applied frequency and the harmonic order increases with the field $A_0$ (see Fig.~\ref{fig:one_band_a-n}c-d). We note that the analytic results are in excellent agreement with the numerical ones (red dotted line) obtained from  Eq.~\ref{eq:group_velocity} in the limit $U=0, V_0=0$. For $V_0\neq 0$, the notion of crystal momentum is no longer valid. However, the qualitative cosine feature of the single band nature of the system is still retained in the intraband physics due to  weak disorder ($V_0<|J|$) except for the emergence of small minigaps as shown in Fig.~(\ref{fig:band_structure}). This is also evident from the intensity spectra presented in Figs.~\ref{fig:HHG_nonint} (a,b).
\begin{figure}
	\centering
	\includegraphics[width=1\columnwidth]{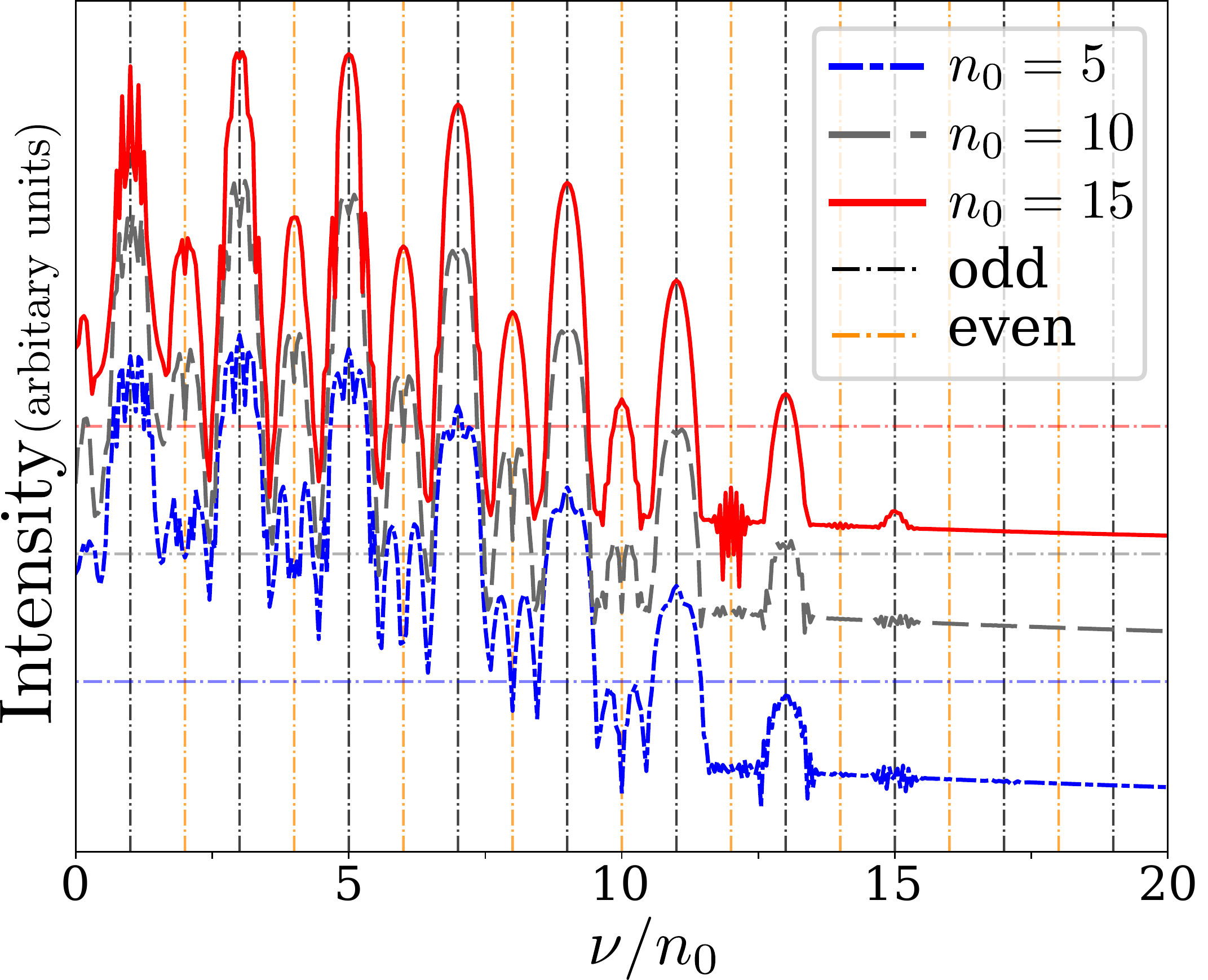}
	\caption{Frequency dependent harmonic order for $n=20$ cycles illustrating the enhancement in the sharpness of the even numbered peaks with the increase in frequency $n_0$ of the applied pulse. The $y-$axis is in arbitrary units. The horizontal lines are the $10^0$ level of respective frequency.}
	\label{fig:freq_hhg}
\end{figure}	
			
	The localized phase with $V_0>2|J|$ is identified by the presence of both even and odd harmonics due to strong breaking of inversion symmetry. The magnitudes of even harmonics are in general subdominant, however can be enhanced by increasing the strength of the field ($A_0$), frequency ($\omega$)  and the number of cycles ($n$) of the applied pulse field. Fig.~(\ref{fig:HHG_nonint})c shows that the even harmonics become more amplified as we increase $A_0$ for a fixed $\omega$ and $n$. However, this feature is limited to the field strength $A_0 \lesssim V_0/J+2$ (see Fig.~(\ref{fig:HHG_nonint})d). Moreover, we find that for a fixed $\omega$ and $A_0$, both the even and odd harmonics become much more prominent on increasing the number of cycles $n$ from 10 to 20 as evident from Fig.~\ref{fig:HHG_nonint}e-g. The origin of this feature lies in the computation of the transition matrix elements between the evolved ground state $|\Psi_0(t)\rangle$ at $(t\neq0)$ with the excited states $|f\rangle$ at the initial time $t=0$. This particularly involves computing $\langle f|H^{int}|\Psi_0(t)\rangle$, where $H^{int}$ is the perturbing Hamiltonian due to the applied pulse. This turns out to be proportional to a Lorentzian in the frequency domain for a finite pulse length, that is, finite $n$. For $n\rightarrow\infty$, the overlap between the ground state at $t\neq0$ with the high-lying states tends to a Dirac delta function peaking at $\nu \omega$, where $\nu$ is an integer. For a detailed calculation we urge the reader to refer to Appendix A.

	Additionally, for a fixed $A_0$ and $n$, the magnitudes of even harmonics can be tuned by varying $\omega$ as shown in Fig.~\ref{fig:freq_hhg}. If the pulse energy ($\hbar\,\omega$) is small compared to the all energy scales in the problem, the magnitudes of even harmonics are found to be negligible. As we increase $\omega$ (equivalently $n_0$), the magnitudes of even harmonics enhance. The reason for such behavior is due to the interband transitions involving minigaps. For small $\omega$, the probability of interband transition is negligible. As we increase $\omega$ the probability of interband transitions increases, and consequently the even harmonics become much more prominent.

	\subsubsection{Field dependent cut-off}
\begin{figure}
\centering
\includegraphics[width=0.95\linewidth]{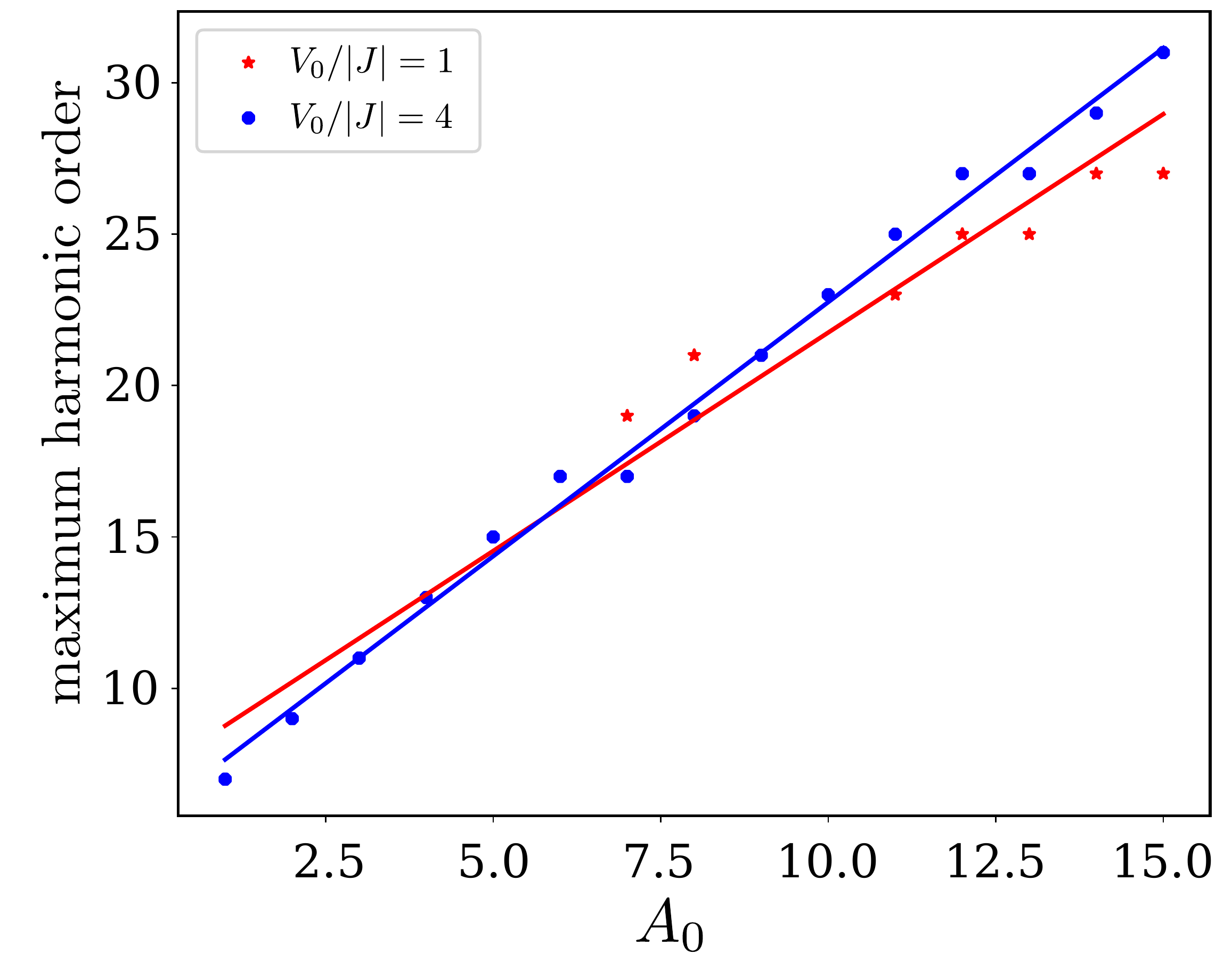}
\caption{Dependence of cut-off with applied field strength in non-interacting region.}
\label{fig:nonint_cutoff}
\end{figure}

	We next compute field dependent cut-off frequency as demonstrated in Fig.~(\ref{fig:nonint_cutoff}). The maximum value of $\nu/n_0$ till which the harmonic peaks appear is called the \emph{cut-off}. In the present scenario, we find the cut-off to have a linear relationship with the strength of the vector potential $A_0$ as evident from Fig.~\ref{fig:nonint_cutoff}. With the increase in $A_0$, the coefficients of Fourier expansion $A_\nu$ of $A(t)=\sum_{\nu}A_{\nu}e^{i\nu\omega t}$ are enhanced. Additionally, with higher $\nu$, the magnitude of the matrix elements denoting transitions between the evolved ground states with the excited states decreases occupying the tails of the Lorentzian and getting deviated away from the central peak value. The interplay of the product of $A_\nu$ with the magnitude of the matrix elements following the Lorentzian governs the increase of the cut-off with $A_0$. That is to say, with higher $A_0$, higher order Fourier coefficient $A_\nu$ begins to contribute towards the appearance of higher order peaks defining the cut-off. The detailed analytical calculation is provided in Appendix A. The introduction of the disordered onsite potential retains the linear dependence of the cut-off on the applied field as shown in Fig.~\ref{fig:nonint_cutoff}. However, the gradient of the same decreases with increasing the strength of the disordered onsite potential $V_0$. This is attributed to the reduced particle current flow in the system with increasing $V_0$ when the particles tend to get more localized. In other words, as the minigaps increases with $V_0$, the probability for interband transition reduces, leading to decrease in cut-off frequency.

	\begin{figure}
	\centering
	\includegraphics[width=0.95\linewidth]{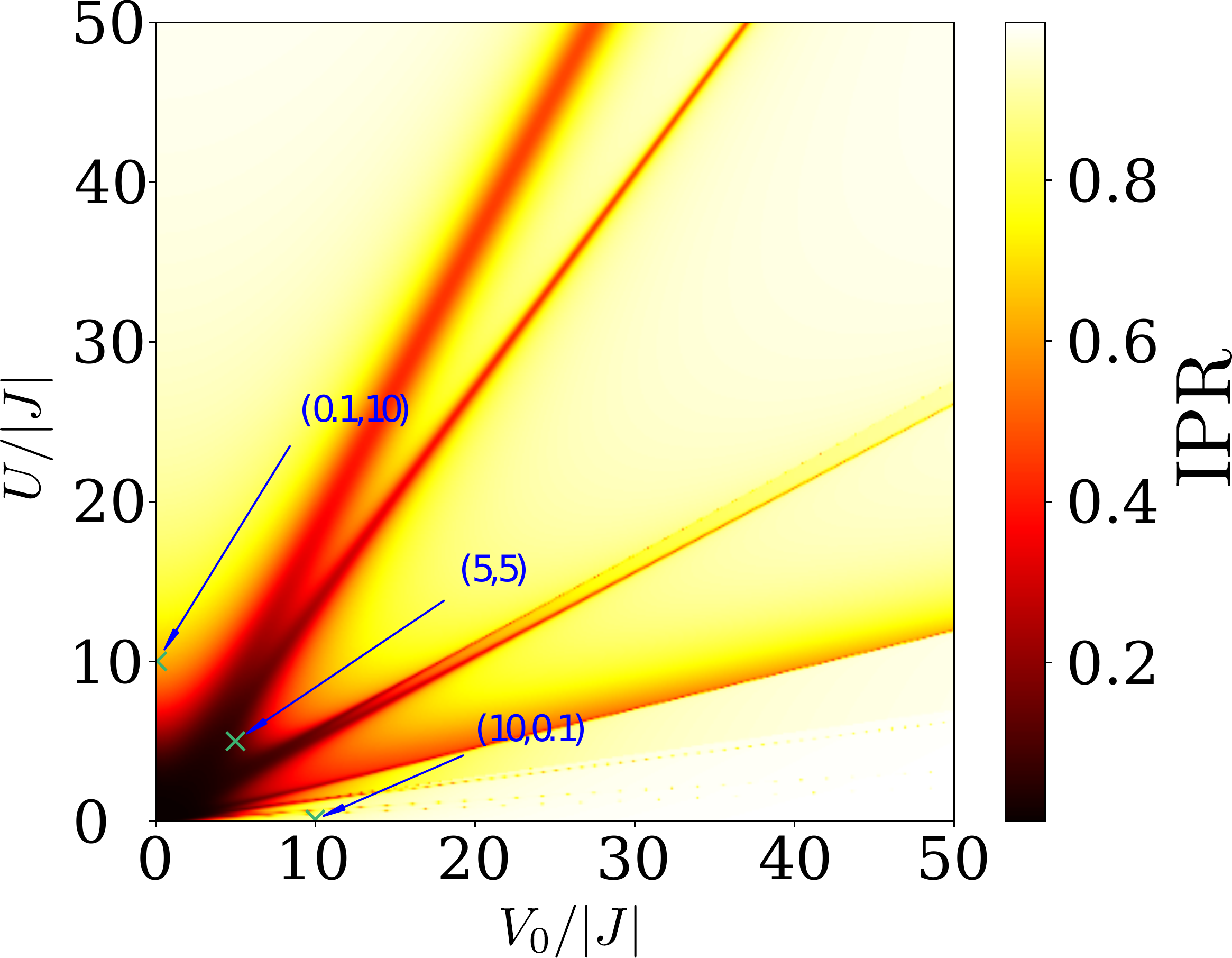}
	\caption{False color-coded image representing the variation of IPR  as a function of interaction ($U/|J|$) and onsite potential ($V_0/|J|)$. The three points mark the three different cases considered in this present work.}
	\label{fig:ipr_3D}
    \end{figure}
		
		\begin{figure*}[htbp]
		\centering
		\includegraphics[width=1.0\textwidth]{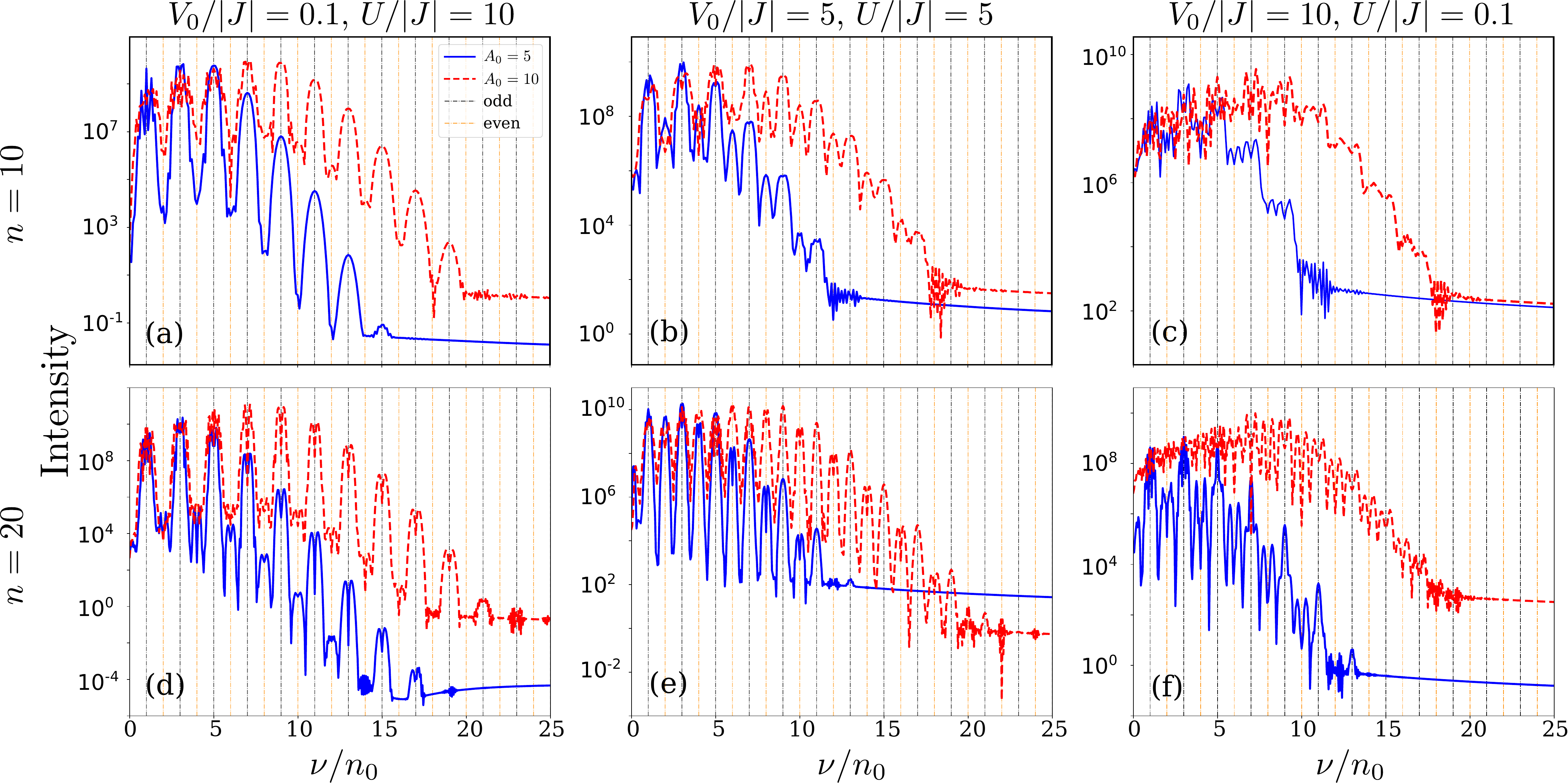}
		\caption{Plots showing intensity spectra with the multiplicity of incident frequency, for number of particles$(N)=7$ and number of lattice sites$(L)=7$ for three representative points marked in the phase diagram Fig \ref{fig:ipr_3D}. (a) $V_0/|J|=0.1$ and $U/|J|=10$, where the IPR $\approx 0.6$. (b) $V_0/|J|=5$ and $U/|J|=5$, where the IPR $\approx 0.2$. (c) $V_0/|J|=10$ and $U/|J|=0.1$, where the IPR $\approx 0.8$ with number of cycles $n=10$; (d), (e) and (f) show the intensity spectra for the aforementioned three points with the number of cycles $n=20$ in the applied pulse.}
		\label{fig:HHG_int}
	\end{figure*}	
		
	\subsection{\label{sec:level10}Interacting case ($U\neq0$)}
	To understand the non-linear response of interacting bosons in the presence of quasiperiodic potential and the underlying mechanism for the generation of harmonic order, we first chart out different phases based on the localization properties. In doing so, we find many body ground state of the interacting Hamiltonian in Eq.~\ref{tidh} using exact diagonalization for system size $L=7$ and particle number $N=7$. This in turn leads to the computation of IPR for different parameters $U$ and $V_0$ for fixed $|J|$. Fig.~\ref{fig:ipr_3D} demonstrates IPR phase diagram in the $V_0/|J|-U/|J|$ plane. We note that the phase diagram obtained from ED for system size with $L=7$ matches qualitatively well with that obtained from DMRG study with bigger system size (say, $L$=35) as shown in Ref.~\onlinecite{giamarchi_PRA08}. Along the $V_0=0$ line the standard Mott insulator and superfluid transition occurs at $U/|J|\equiv 4$ as the IPR is around 25. Along the $U=0$ line disorder driven localization-delocalization transition occurs near $V_0/|J|\equiv 2$, corroborating the phases obtained in the non-interacting case discussed in the preceding section. For finite $V_0$ and $U$, we obtain re-entrant localized and delocalised phases depending on the values of $V_0/|J|$ and $U/|J|$ as evident from Fig.~(\ref{fig:ipr_3D}).  The localization due to interaction turns out to differ from the disorder-induced localization as the configurations of particle distribution differs. At $t=0$, the particle distribution is obtained using the square modulus of the coefficient ($|c_{n_{\alpha_1}n_{\alpha_2}.....n_{\alpha_p}...}(0)|^2$) of individual many-particle basis states of the ground state wavefunction $|\Psi_0(t)\rangle=\sum_{n_{\alpha_1}n_{\alpha_2}..n_{\alpha_p}..}c_{n_{\alpha_1}n_{\alpha_2}..n_{\alpha_p}..}|n_{\alpha_1}n_{\alpha_2}..n_{\alpha_p}...\rangle$, where $|n_{\alpha_1}n_{\alpha_2}..n_{\alpha_p}...\rangle$ denotes normalized state with $n_{\alpha_1}$ particles in state $|\alpha_1\rangle$, $n_{\alpha_2}$ particles in state $|\alpha_2\rangle$,$\cdots$ and $\{|\alpha_i\rangle\}$ is an orthonormal basis.  For high values of $V_0/|J|\gg U/|J|$, the particles tend to accumulate in a particular site. In contrast, for $V_0/|J|\ll U/|J|$, the particles tend to be distributed equally in each site with equal density, leading to the typical Mott localization. In Table \ref{tab:table1}, we provide probable particle distributions in lattice sites for different values of $U/|J|$ and $V_0/|J|$.

	\begin{center}
		\begin{table}[H]
			\caption{\label{tab:table1} Most probable particle configuration in different values of $V_0/|J|$ and $U/|J|$}
			\begin{ruledtabular}
				\begin{tabular}{||c c c||} 
					\hline
					$V_0/|J|$ & $U/|J|$ & most probable configuration \\ [0.5ex] 
					\hline\hline
					10 & 0.1 & [0~0~0~7~0~0~0] \\ 
					\hline
					5 & 5 & [2~1~0~2~0~1~1] \\
					\hline
					0.1 & 10 & [1~1~1~1~1~1~1] \\
					\hline
				\end{tabular}
			\end{ruledtabular}
		\end{table}
	\end{center}
	
		\begin{figure*}
		\centering
		\includegraphics[width=\textwidth]{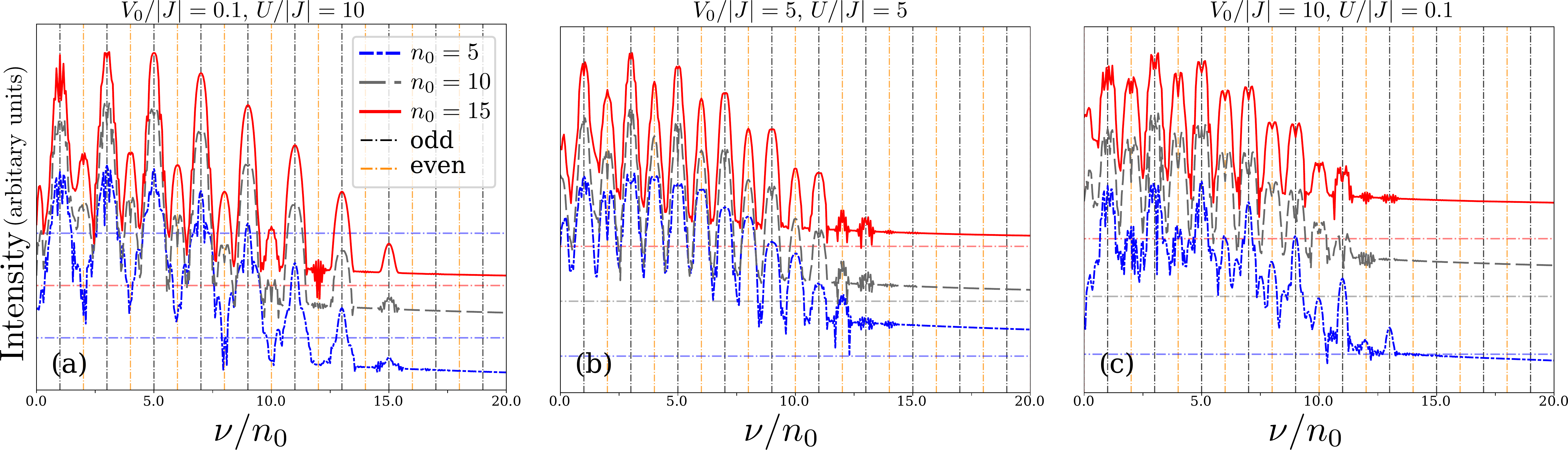}
		\caption{Frequency dependent harmonic order for $n=20$ cycles. Similar to Fig.~(\ref{fig:freq_hhg}), the plots are shifted by arbitrary $y-$ values for visual aid.  The horizontal lines denote $10^0$ value for their respective colors. (a) $V_0/|J|=0.1$, $U/|J|=10$, (b) $V_0/|J|=5$, $U/|J|=5$, (c) $V_0/|J|=10$, $U/|J|=0.1$.}
		\label{fig:freq_inter}
	\end{figure*}
\begin{figure}[b]
	\centering
	\includegraphics[width=0.99\linewidth]{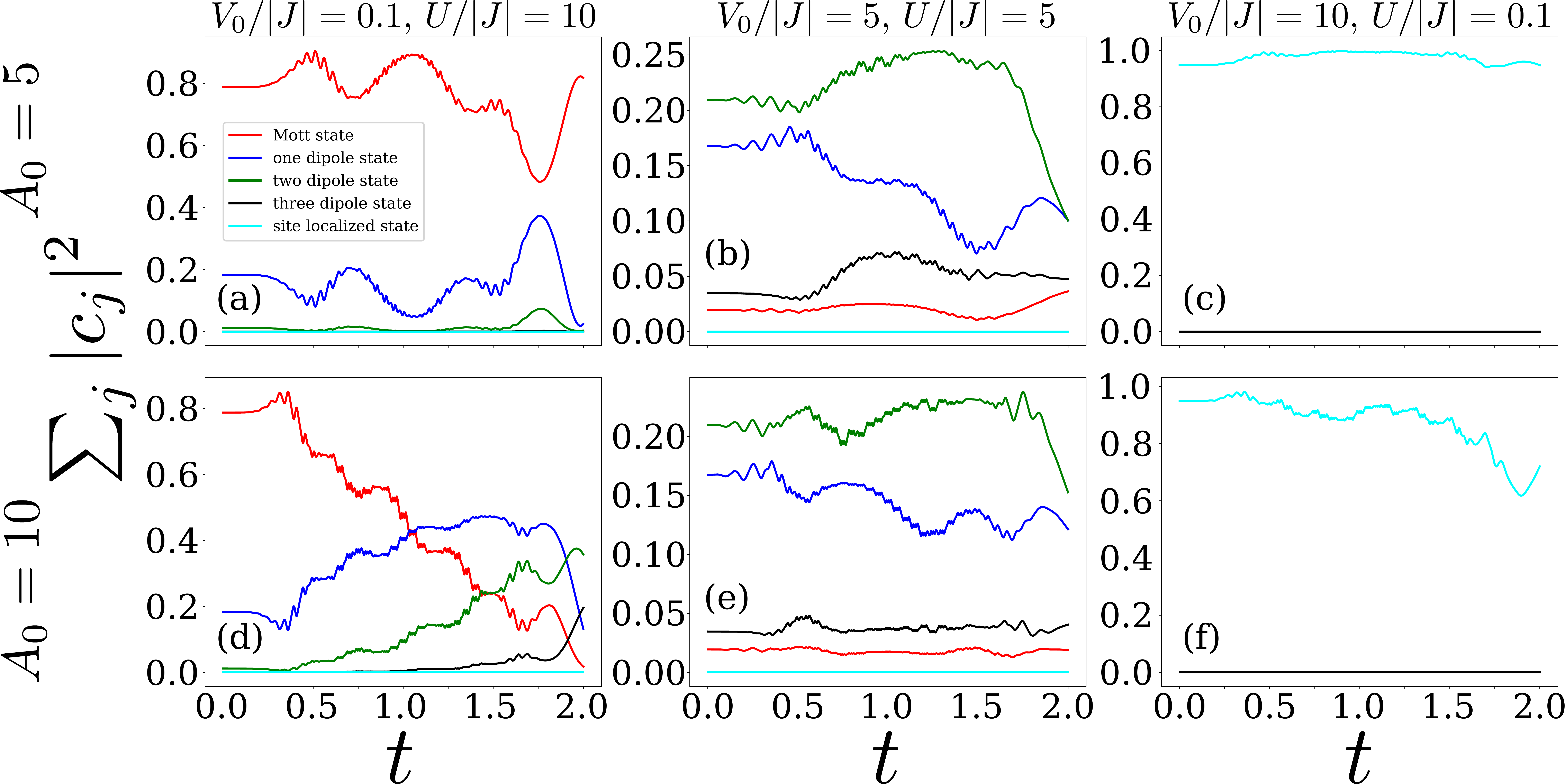}
	\caption{Contribution of particular state $j\in \{{\rm Mott}~ {\rm or}~ 1-{\rm  or}~ 2-{\rm or}~ 3-{\rm dipole}~{\rm or}~ {\rm site~ localized}\}$ states in the ensuing dynamics towards the generation of higher harmonics. Figure (a), (b) and (c) illustrate the dynamics for ($V_0/|J|=0.1$, $U/|J|=10$), ($V_0/|J|=5$, $U/|J|=5$) and ($V_0/|J|=10$, $U/|J|=0.1$) respectively with $A_0=5$. Figure (d), (e) and (f) show the dynamics with similar parameters for $A_0=10$. Here $t$ is measured in picoseconds.}
	\label{fig:dipole_contribution}
\end{figure}

	\subsubsection{Intensity spectra and mechanism}
	Having discussed the possible phases, we now focus on the response of both interaction-driven localization and disorder-driven localization to the pulse field. Fig.~\ref{fig:HHG_int} represents the intensity spectra for the three representative regimes based on the probable particle configurations in Table  \ref{tab:table1}. Let us first focus on the $U/|J|\gg V_0/|J|$ limit (see Fig.~(\ref{fig:ipr_3D})), where particles are distributed equally in each lattice sites. For a fixed $\omega$ and $n$,  the interaction-driven localized phase contains only odd harmonics (Fig.~\ref{fig:HHG_int}a) similar to the case of delocalized phase of non-interacting Hamiltonian (Fig.~\ref{fig:HHG_nonint}a). Interestingly, the even harmonics may emerge in this interacting regime if we vary $\omega$ and $n$.   
	Fig.~(\ref{fig:freq_inter}) demonstrates this feature. The increase in $\omega$ indeed facilitates the substantial interband transitions for even orders within the Mott gap and quasiperiodicity-induced minigaps. The interaction however alone cannot produce even harmonics irrespective of the variation in $n$ and $\omega$ because of the presence of inversion symmetry. This is one of the key findings of the present paper. 
	
	With $U/|J|\ll V_0/|J|$ fixing $U_0/|J|=0.1$, the localization is mainly governed by the quasi-periodicity as measured through IPR given in Fig.~\ref{fig:ipr_3D}, where all the particles are localized in a single site.  In this case, we do not see any additional feature in the intensity pattern when compared to the completely non-interacting $(U=0)$ localised phase (see Fig.~\ref{fig:HHG_int}c and Fig.~\ref{fig:freq_inter}c). 
	For $U/|J|\sim V_0/|J|<10$,  the system is in delocalised phase as evident from Fig.~\ref{fig:ipr_3D}. The delocalised phase in the presence of interactions seems to respond differently than the limiting cases discussed in the preceding paragraphs. In this case, even for $n=10$-cycle pulse, we obtain comparable even and odd harmonics as evident from Fig.~(\ref{fig:HHG_int}b). As we increase the field strength, the harmonic order is enhanced. Moreover, both even and odd peaks become more prominent if we increase number of pulse cycle to $n=20$ as clearly shown in Fig.~\ref{fig:freq_inter}b. Thus the delocalized phase with approximately equal interaction and disorder strengths presents a completely new feature in the harmonic spectra when contrasted with the other scenario. This is another important and interesting result obtained in the present model. In the next paragraph, we investigate the role of the excited states that are responsible for giving rise to harmonic orders in different parameter regimes.
	
	To understand the presence of harmonic orders in the current spectrum, we identify the evolved excited states that are primarily responsible for the current to contain multiple frequencies of the applied field. For $U/|J|\gg V_0/|J|$, the dynamics is governed by the Mott ground state (e.g., $|11111111\rangle$) accompanied by the contribution from all the possible excited \emph{single dipole} states~\cite{subir_PRB02} where a quasiparticle-quasihole pair resides on nearest-neighbour sites such as $|1021111\rangle$ (see Fig.~\ref{fig:dipole_contribution}a,d).  When $U/|J|=V_0/|J|=5$, that is in the interacting delocalised phase, the contribution to current is mainly governed by the formation of single and \emph{two dipole} states (e.g., $|1020211\rangle$); while the contribution from all other possible states are suppressed (see Fig.~\ref{fig:dipole_contribution}b,e). On the other hand, in the deep localized phase with $V_0/|J|\gg U/|J|$ and single site occupancy, the site-localized state ( non-resonant state) gives rise to higher harmonics as shown in Fig.~(\ref{fig:dipole_contribution}c,f).
	
\begin{figure}[t]
		\centering
		\includegraphics[width=0.95\linewidth]{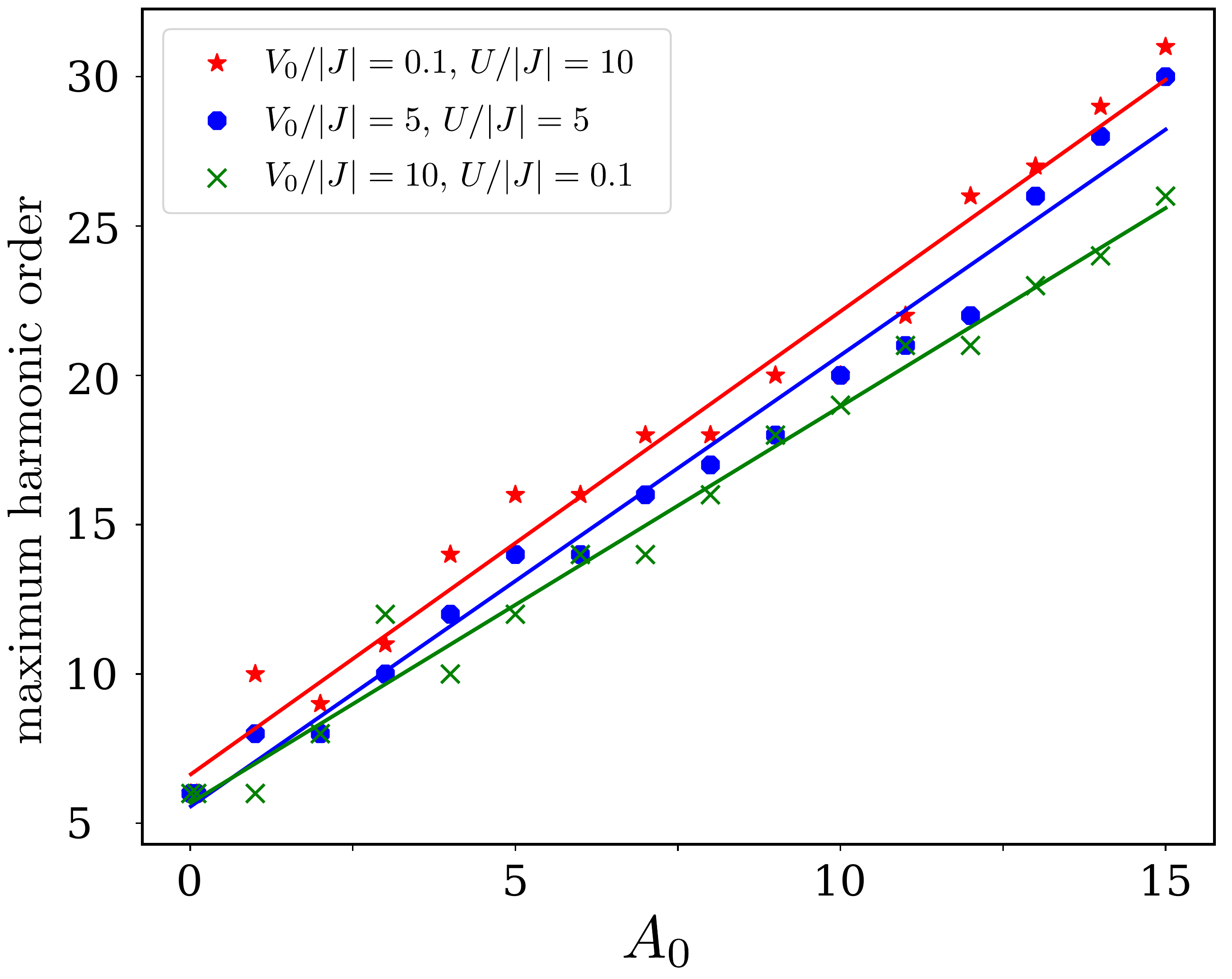}
		\caption{Dependence of cut-off with applied field strength in interacting region.}
		\label{fig:int_cutoff}
	\end{figure}

	Finally, we show in Fig.~\ref{fig:int_cutoff} the field dependent cutoff for all representative parameter regimes discussed above. It turns out that the interaction does not affect the linear dependence as obtained for non-interacting case.


	\section{Conclusion}
	We investigate the non-linear response of interacting bosonic model to an electric pulse field in the presence of an incommensurate potential. We find that the quasiperiodicity driven localized-delocalized phases respond differently to the electric pulse field in the presence and absence of interaction. While the delocalised phase does not exhibit odd harmonics, the localised phase can contain both even and odd harmonics and the amplitudes of odd and even harmonics can be enhanced or reduced by varying the frequency of the applied field. Moreover, the cycles of the pulse can be used to sharpen the peaks of harmonic orders. In the interacting case, we obtain even richer physics as we tune frequency, cycles, and amplitudes of the pulse. For reasonably large interaction compared to the disorder potential, the localised Mott phase can exhibit both even and odd harmonics depending on the frequency, field strength and number of pulse cycles. This is in contrast to the interacting phase without any disorder. Further, for comparable disorder and interaction strength, the delocalised phase remarkably shows even and odd harmonics with equal magnitudes. This fact can be used as a key to distinguish noninteracting delocalised from that of interacting delocalised phases. In addition, while the system is completely localised induced primarily by disorder with small interaction, the even and odd harmonics can be obtained similar to non-interacting localised phases. However, the presence of a weak interaction can give rise to comparable even and odd harmonics with increasing frequency. Thus the interplay between interaction and disorder play important role in generating both even and odd harmonics with comparable magnitudes in the present study. To this end, we note that such interacting model together with additional quasiperiodic potential can be easily attainable in cold atomic settings. Moreover, creating an synthetic electric field is a routine now a days. Thus our theoretical results can easily be verified in experiments. 	
	\section{Acknowledgement}
	KS thanks Anamitra Mukherjee for useful discussion. DD acknowledges use of Virgo cluster at NISER.

	\clearpage
\begin{widetext}	
	\appendix
		\section{\label{sec:level5}Two level Model and pulse-driven transitions}
		
		In this section, we provide an approximate analytic expression for transition amplitudes between two states of a generic Hamiltonian in the presence of an external electromagnetic field. This will allow us to understand the presence of harmonic orders in the current discussed in the main text. We start with the Hamiltonian 
		\be
		\hat{H}_0=\sum_i\frac{\vec{p}_i^2}{2m}+\sum_iV(r_i).
		\label{appn:ham0}
		\ee
		
		In the presence of an external electromagnetic field, Eq.~(\ref{appn:ham0}) can be written as
		\be
		\hat{H}=\sum_i\frac{(\vec{p}_i-q\vec{A}(\vec{r},t))^2}{2m}+\sum_iV(\vec{r}_i)+q\phi(\vec{r}_i,t)
		\ee
		where the vector potential $\vec{A}$ and scalar potential $\phi(\vec r,t)$ can be obtained via $\vec{E}=-\frac{\partial \vec{A}(\vec{r},t)}{\partial t}$ and $\vec{E}=-\vec{\nabla}\phi(\vec{r},t)$, respectively. In velocity gauge, the Hamiltonian can be rewritten as 
		
		\bea
		\hat{H}_v&=\sum_i\frac{(\vec{p}_i-q\vec{A}(\vec{r}_i,t))^2}{2m}+\sum_iV(\vec{r}_i)=H_0+\hat{H}^{int}_v \label{eq:velocity_gauge}
		\eea
		where
		\be
		H^{int}_v=\frac{1}{2m}\sum_i(qA(r_i,t).p_i+qp_i.A(r_i,t)+q^2A^2(r_i,t))
		\ee
		Using Coulomb gauge $\nabla.\vec{A}=0$ and keeping only the linear order of field strength, we find 
		\bea
		H^{int}_v&=&\frac{2q}{2m}\sum_iA(r_j,t).p_i=-i\frac{q}{\hbar}\sum_jA(r_j,t)[r_j,H_0]
		\eea
		With this, we compute transition amplitude in the interaction picture using $\langle f|U_I(t,t_0)|i\rangle$, where $U_I(t,t_0)=\mathcal{I}+\sum_jU_I^{(j)}(t,t_0)$ and 
		\bea
		U_I^{(1)}(t,t_0)&=&-\frac{i}{\hbar}\int_{t_0}^{t}V_I(t')dt'\\
		U_I^{(2)}(t,t_0)&=&-\frac{i}{\hbar}\int_{t_0}^{t}V_I(t_1)dt_1\int_{t_0}^{t_1}V_I(t_2)dt_2\eea
		where \bea
		V_I(t)&=&e^{i H_0 t/\hbar}H^{int}_ve^{-i H_0 t/\hbar}
		\eea
		 For $A(r,t)=A(t)=\sum_{\nu}A_{\nu}e^{-\iota \nu\omega t}$, we obtain
		 \bea
		\langle f|U_I^{(1)}(t,t_0)|i\rangle&=&-\frac{i}{\hbar}\int_{t_0}^{t}\langle f|V_I(t')|i\rangle dt'\nonumber\\
		&=&-\frac{q}{\hbar^2}\int_{t_0}^{t}\sum_{\nu}A_{\nu}e^{-\iota \nu\omega t'}e^{i (\omega_f-\omega_i)t'}\langle f|\sum_i[r_i,H_0]|i\rangle dt'\nonumber\\
		&=&-\frac{q}{\hbar^2}\int_{t_0}^{t}\sum_{i,\nu}A_{\nu}e^{-\iota \nu\omega t'}e^{i (\omega_f-\omega_i)t'}\langle f|(r_i.H_0-H_0.r_i)|i\rangle dt'\nonumber\\
		&=&-\frac{1}{\hbar}\int_{t_0}^{t}\sum_{\nu}A_{\nu}e^{-\iota \nu\omega t'}e^{i (\omega_f-\omega_i)t'}(\omega_f-\omega_i)\langle f|\sum_iqr_i|i\rangle dt'\nonumber\\
		&=&-\frac{1}{\hbar}\int_{t_0}^{t}\sum_{\nu}A_{\nu}e^{-\iota \nu\omega t'}e^{i (\omega_f-\omega_i)t'}(\omega_f-\omega_i)\langle f|D|i\rangle dt'\label{eq:17}
		\eea
		If we consider that the system was in state $|i\rangle$ in deep past, i.e. $t_0\to -\infty$ and we switch off the perturbation in far future, i.e. $t\to\infty$ compared to the dynamics of the system, the  equation \ref{eq:17} can be  recasted as
		\bea
		\langle f|U_I^{(1)}(t,t_0)|i\rangle
		=-\frac{1}{\hbar}\sum_{\nu}A_{\nu}\delta(\omega_f-\omega_i-\nu\omega)(\omega_f-\omega_i)\langle f|D|i\rangle\label{eq:18}
		\eea
		Similarly,
		\bea
		&&\langle f|U_I^{(2)}(t,t_0)|i\rangle\nonumber\\
		&=&(-\frac{i}{\hbar})^2\langle f|\int_{t_0}^{t}e^{i H_0  t_1/\hbar}H^{int}_v(t_1)_ve^{-i H_0  t_1/\hbar}d t_1\int_{t_0}^{t_1}e^{i H_0  t_2/\hbar}H^{int}_v(t_2)e^{-i H_0  t_2/\hbar}|i\rangle d t_2\nonumber\\
		&=&(-\frac{i}{\hbar})^2\sum_j\int_{t_0}^{t}e^{i \omega_f  t_1}\sum_{\nu}A_{\nu}e^{-i \nu\omega t_1}\langle f|D|j\rangle e^{-i \omega_j  t_1}d t_1(\omega_f-\omega_j)(\omega_j-\omega_i)\int_{t_0}^{t_1}e^{\iota \omega_j  t_2}\sum_{\nu'}A_{\nu'}e^{-\iota \nu'\omega t_2}\langle j|D|i\rangle e^{- \omega_i  t_2}d t_2\nonumber\\
		&=&(-\frac{i}{\hbar})^2\sum_j\sum_{\nu}\int_{t_0}^{t}A_{\nu}e^{i \omega_f  t_1-i \nu\omega t_1-i \omega_j  t_1}\langle f|D|j\rangle d t_1(\omega_f-\omega_j)(\omega_j-\omega_i)\sum_{\nu'}\int_{t_0}^{t_1}A_{\nu'} e^{i \omega_j  t_2-i \omega_i  t_2-i \nu'\omega t_2}\cross\langle j|D|i\rangle d t_2\nonumber\\
		&=&(-\frac{i}{\hbar})^2\sum_j\sum_{\nu}\int_{t_0}^{t}A_{\nu}e^{i \omega_f  t_1-i \nu\omega t_1-\iota \omega_j  t_1}\langle f|D|j\rangle d t_1(\omega_f-\omega_j)(\omega_j-\omega_i)\sum_{\nu'}A_{\nu'} \frac{e^{\i \omega_j  t_2-\iota \omega_i  t_2-i \nu'\omega t_2}}{i \omega_j-i \omega_i-\iota \nu'\omega}\bigg|_{t_0}^{t_1}\langle j|D|i\rangle\nonumber\\
		&=&(-\frac{i}{\hbar})^2\sum_{j,\nu,\nu'}\int_{t_0}^{t}A_{\nu}A_{\nu'} \frac{e^{i \omega_f  t_1-\iota \omega_i  t_1-i \nu\omega t_1-i \nu'\omega t_1}}{\iota \omega_j-i \omega_i-i \nu'\omega}dt_1(\omega_f-\omega_j)(\omega_j-\omega_i)\langle f|D|j\rangle\langle j|D|i\rangle\nonumber\\
		&=&(-\frac{i}{\hbar})^2\sum_{j,\nu,\nu'}\int_{t_0}^{t}A_{\nu}A_{\nu'} \frac{e^{i \omega_f  t_1-\iota \omega_i  t_1-\iota \nu\omega t_1-i \nu'\omega t_1}}{\iota \omega_j-i \omega_i-i \nu'\omega}dt_1(\omega_f-\omega_j)(\omega_j-\omega_i)\langle f|D|j\rangle\langle j|D|i\rangle\nonumber\\
		&=&(-\frac{i}{\hbar})^2\sum_{j,\nu,\nu'}A_{\nu}A_{\nu'} \frac{\delta(\omega_f -\omega_i - (\nu+\nu')\omega)}{\iota \omega_j-i \omega_i-i \nu'\omega}(\omega_f-\omega_j)(\omega_j-\omega_i)\langle f|D|j\rangle\langle j|D|i\rangle.
		\eea
		In a similar way, it is easy to find $\langle f|U_I^{(3)}(t,t_0)|i\rangle$ contains  $\delta(\omega_f-\omega_i-(\nu+\nu'+\nu'')\omega)$ and higher orders as well. If the external field contain only one Fourier component, e.g., $\nu=1$, then the transition amplitude is non-zero only when $\omega_f-\omega_i=\omega$, $\omega_f-\omega_i=2\omega$, $\omega_f-\omega_i=3\omega$ and so on. 
		Thus for the incident light with lower frequency than the energy difference between the two eigenstates, the transition amplitude, $P=|\langle f|U_I(t,t_0)|i\rangle|^2$ can contain multiple frequencies of the incident light, provided that the process should be adiabatic or slow enough compared to the dynamics of the system. The process is typically called multiphoton process.

\end{widetext}	
	
	\bibliographystyle{apsrev4-1}
	\bibliography{references}
	
\end{document}